\documentstyle[mprocl]{article}

\def\Journal#1#2#3#4{{#1} {\bf #2}, #3 (#4)}

\def\be{\begin{equation}}
\def\ee{\end{equation}}
\def\bea{\begin{eqnarray}}
\def\eea{\end{eqnarray}}

\begin{document}

\title{PAIR OF ACCELERATED FRAMES: 
\marginpar{\emph{Based on the second of
two talks given June 23-24, 1997, at the 8th Marcel Grossmann Meeting,
held in Jerusalem, Israel.}}
A PERFECT INTERFEROMETER}

\author{ ULRICH H. GERLACH }

\address{Department of Mathematics, Ohio State University, Columbus,
OH 43210, USA}


\maketitle\abstracts{
The four Rindler quadrants of a pair of oppositely accelerated frames
are identified as a (Lorentzian) Mach-Zehnder interferometer.
The Rindler frequency dependence of the interference process is
expressed by means of a (Lorentzian) differential cross section. The
Rindler frequencies of the waves in the two acccelerated frames can be 
measured directly by means of a simple inertially moving detector.
}

Does there exist a purely quantum mechanical carrier of the imprints
of gravitation? Mathematical aspects as well as the motivation for
considering this question are discussed elsewhere in these proceedings.
Here we summarizes some key physical aspects
of this question.

Gravitation manifests itself by the imprints it leaves on the geodesic
world lines of particles. It is a remarkable fact that even though
particles may differ from one another in their mass and charge, the
manner in which they reveal the presence of gravitation is totally
independent of these intrinsic properties. It is stricly by means of
their classical paths in spacetime that particles capture the imprints
of gravitation. The mass and charge, i.e. internal composition, of
these particles plays no role (uniqueness of free fall, ``weak
equivalence principle'').

The last property is a key requirement dictated by the Dicke-Eotvos
experiment. Adopting its extension to quantum mechanics (``imprints of
gravitation are independent of a particle's Compton wave length and
kinetic energy'') has led us to pairs of oppositely accelerating
(``Rindler'') frames as the spacetime framework for measuring the
presence of gravitation. 

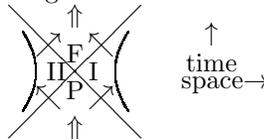
\begin{picture}(100,50)(-100,0)
\put(32,25){\makebox(0,0){I}}
\put(18,25){\makebox(0,0){II}}
\put(25,32){\makebox(0,0){F}}
\put(25,18){\makebox(0,0){P}}
\put(15,15){\makebox(0,0){$\nwarrow$}}
\put(35,35){\makebox(0,0){$\nwarrow$}}
\put(15,35){\makebox(0,0){$\nearrow$}}
\put(35,15){\makebox(0,0){$\nearrow$}}
\put(25,4){\makebox(0,0){$\Uparrow$}}
\put(25,46){\makebox(0,0){$\Uparrow$}}
\put(82,20){\makebox(0,0){space$\rightarrow$}}
\put(67,25){\shortstack{$\uparrow$\\time}}
\qbezier(45,10)(35,25)(45,40)
\qbezier(5,10)(15,25)(5,40)
\put(0,0){\line(1,1){50}}
\put(0,50){\line(1,-1){50}}
\put(25,25){\circle*{1}}
\end{picture}

More precisely, the two accelerating frames serve as the two coherent
legs of a Lorentzian version of the Mach-Zehnder interferometer. The
two well-known pseudo-gravitational potentials in these frames serve as the two
mirrors with 100\% reflectivity. The two spacetime regions in the
future $F$ and the past $P$ \emph{near} the intersection
(``bifurcation event'') of the event horizons serve as
``half-silvered'' mirrors. A wave in $P$ \emph{far} from the
bifurcation event enters the ``interferometer'' from $P$.  
One can show mathematically that near the
bifurcation event the wave splits into two partial waves: one
propagates across the past event horizon, enters Rindler Sector $I$,
and gets reflected by its potential, the other does the analogous
thing in Rindler Sector $II$.  The two partial waves recombine
\emph{near} the bifurcation event in $F$ and then leave the
``interferometer'' by proceeding towards a detector situated
\emph{far} from the bifurcation event.  The detector (in $F$) is
inertial, and it measures the Minkowski particle (or anti-particle)
number.

The detector will record an interference pattern as follows: into
accelerating Rindler frame $I$ place a dielectric slab so that it is static in
(i.e. coaccelerating with) that frame. Because the refractive index
differs from unity, the wave reflected in that frame will have
suffered a phase shift. This phase shift depends on the thickness of
the slab.  Upon combining with the wave reflected in Rindler frame $II$,
the phase shifted wave produces an alteration in the Minkowski
particle count recorded by the detector in $F$. The interference
pattern itself consists of the recorded particle count as a function
of the thickness of the dielectric slab.

The determination of the refractive index with measurements based on an
interference pattern illustrates the versatility of the ``Lorentizian''
Mach-Zehnder interferometer. Its two ``legs'' (Rindler Sectors $I$ and $II$)
\noindent not only (i) furnish a spacetime environment with a Cauchy
hypersurface for the propagation of disturbaces governed by a wave
equation,
\noindent nor do they (ii) only serve as a nature-given arrangement for
determining the effective parameters which characterize the
propagation environment,
\noindent but (iii) they also have a pair of effective
(``pseudo-gravitational'') potentials which provide a \emph{diffractive
aperture} through which waves diffract as they propagate from the
past ($P$) to the future ($F$).


The interference pattern of a Lorentzian Mach-Zehnder interferometer is
based on globally defined waves which are monochomatic, i.e.
Lorentz boost invariant. One would
have to consider such waves if the above ``versatility'' of the
interferometer is to be realized. These wave modes have been
exhibited explicitly [1], and they all satisfy the
same wave equation, namely ${\partial^2 \psi \over \partial t^2}-
{\partial^2 \psi \over \partial z^2}+ k^2\psi =0$, where $k^2=k^2_x
+k^2_y +m^2 $, or
\begin{equation}
\left[\frac{1}{\xi ^2} \frac{\partial^2}{\partial\tau^2} -\frac{1}{\xi}
\frac{\partial}{\partial\xi} \xi \frac{\partial}{\partial\xi}
+k^2\right]\psi =0~~\hbox{in}~I ~\hbox{and}~II
\end{equation}
and
\begin{equation}
\left[-\frac{1}{\xi ^2} \frac{\partial^2}{\partial\tau^2} +\frac{1}{\xi}
\frac{\partial}{\partial\xi} \xi \frac{\partial}{\partial\xi}
+k^2\right]\psi =0~~\hbox{in}~P ~\hbox{and}~F.
\end{equation}
Even though the propagation of each wave mode from $P$ to $F$
constitutes a unique and separate Mach-Zehnder interference process,
the nature-given interferometer for these processes consists of one
and the same pair of Rindler frames, $I$ and $II$. The only difference
is the obvious one: the effective location ($\xi$) of the 100\%
reflective mirrors depends on the boost frequency of these modes.
In fact, the effective proper separation between these two mirrors
is $2\times (Rindler~frequency)/k$ for a given mode. Beyond this distance
the wave becomes evanescent.

Consider the propagation of a positive Minkowski frequency plane wave
mode from $P$ to $F$.  It is a superposition of the globally defined
monochromatic (i.e. pure Rindler frequency) wave modes, each one
propagating from $P$ to $F$ and hence characterized by its own
Mach-Zehnder interference process.  This means that the propagation of
a plane wave mode is equivalent to the simultaneous occurrence of all
the corresponding Mach-Zehnder interference processes. There is one such
process for each Rindler frequency. The
interference is extremely delicate and of a very special kind.  In $F$,
each pair of monochromatic partial waves (having the same
\emph{spatial} Rindler frequency and coming from $I$ and $II$
respectively) has a phase relation such that an inertial detector,
sensitive to the spatial frequency of these waves, will measure only
particles, and no antiparticles.

Suppose there is a gravitational (or some other) disturbance in one or
both of the two coherent legs of the interferometer. Such a disturbance
will spoil the delicate interference.  If the disturbance is Lorentz
invariant, then the interference will be altered in a very simple and
special way: only the phases in each of the pair of partial monochromatic wave
modes from $I$ and $II$ will be altered. No mixing between waves of different 
Rindler frequency. Let $\delta_I(\omega)$ and $\delta_{II}(\omega)$
be the phase shift of the two wave amplitudes reflected from $I$ and $II$.
Then the \emph{partial wave cross section} for the scattering of a 
plane wave by the Lorentz invariant disturbance is [2]
\begin{equation}
\frac{d\sigma}{d\omega}
=\frac{1}{k\cosh\theta}~\big\{1+\big( \frac{\sin(\delta_I(\omega)-\delta_{II}
(\omega))}{\sinh\pi\omega}\big)^2\big\}
\end{equation}
Here $k\cosh \theta$ is the Minkowski frequency of the plane wave in
$P$ before it got scattered.  This partial wave cross section
expresses the interference pattern mentioned at the top of the
previous page.

If the gravitational disturbance is \emph{not} Lorentz invariant
relative to the pair of accelerated frames, then the Mach-Zehnder
interference processes will \emph{not} be independent of one another,
and there will be a mixing between the different Rindler frequencies.

Regardless of the detailed nature of the disturbance, the task of
identifying it by means of its particle (or antiparticle) spectrum in 
$F$ falls on the set of Rindler frequency selective particle detectors
moving inertially in $F$. These detectors are discussed below.
The information which they record is, roughly speaking, the intensity of the
(Rindler frequency) fourier transform of the disturbance.

The physical justification for identifying the four Rindler sectors as
a Lorentzian Mach-Zehnder interferometer depends on measuring the
interfering Rindler frequency components from Sectors $I$ and $II$.
The waves having a specific Rindler frequency propagate across the
future event horizon into Rindler sector $F$. There each of these
waves is selectively measured by an \emph{inertially moving} detector
with a Fabry-Perot interference filter whose mirror separation
increases with constant speed. Such a filter-detector combination in
$F$ responds only to \emph{discrete Rindler frequencies}
\begin{equation}
\omega=\frac{\pi \ell}{\tanh ^{-1} \beta}
\quad , \quad  \ell =1,2,\cdots
\end{equation}
Here $\beta$ is the relative speed of the two mirrors. These frequencies are 
the transmission resonances of the expanding Fabry-Perot cavity. The 
sharpness of these resonances, and hence the selectivity of the interference
filter, depends on the reflectivity $(<1)$ of the Fabry-Perot mirrors. Waves
with frequencies different from the discrete resonances get reflected, and 
hence do not interact with the detector.

Waves from Rindler $I$ enter the cavity through one mirror, while those
from Rindler $II$ enter through the other. Coupled to the waves trapped
between these two nearly, but not quite, completely silvered mirrors,
the inertial detector lends itself to recording the Lorentzian
Mach-Zehnder interference process.

\noindent \textbf{References}
\par
[1] U.H. Gerlach, Phys. Rev. D 38, 514 (1988)
\par
[2] U.H.Gerlach, ``Scattering by a Pair of Oppositely Accelerated Dielectric
Media'', preprint (1997)

\end{document}